# Network Architecture Design toward Convergence of Mobile Applications and Networks

Shuangfeng Han, Zhiming Liu,Tao Sun, and Xiaoyun Wang (Corresponding Author)
China Mobile Research Institute

***Abstract*:** With the quick proliferation of extended reality (XR) services, the mobile communications networks are faced with gigantic challenges to meet the diversified and challenging service requirements. A tight coordination or even convergence of applications and mobile networks is highly motivated. In this paper, a multi-domain (e.g. application layer, transport layer, the core network, radio access network, user equipment) coordination scheme is first proposed, which facilitates a tight coordination between applications and networks based on the current 5G networks. Toward the convergence of applications and networks, a network architectures with cross-domain joint processing capability is further proposed for 6G mobile communications and beyond. Both designs are able to provide more accurate information of the quality of experience (QoE) and quality of service (QoS), thus paving the path for the joint optimization of applications and networks. The benefits of the QoE assisted scheduling are further investigated via simulations. A new QoE-oriented fairness metric is further proposed, which is capable of ensuring better fairness when different services are scheduled. Future research directions and their standardization impacts are also identified. Toward optimized end-to-end service provision, the paradigm shift from loosely coupled to converged design of applications and wireless communication networks is indispensable.

*Index terms*-Quality of experience, quality of service, network architecture, fairness index, scheduler, 6G

## I. INTRODUCTION

One of the fundamental motivations for the design and operation of the wireless communication networks is to ensure the end-to-end mobile service requirements are satisfied efficiently. However, this is actually a challenging task. Firstly, there has long been a loose coordination between the over-the-top (OTT) companies and the mobile operators. The quality of service (QoS) [1] and quality of experience (QoE) [2,3] of the mobile services, which are of paramount significance to the OTTs, are in reality controlled and managed by the operators. The OTTs are unwilling to pay additional money to the operators for enhanced guarantee of the QoS and QoE. Rather, they would like to see updated versions of specifications in the standard bodies to make sure their services will be better supported in the wireless networks. For information security and privacy purpose, the OTTs may want to cipher their services and would not allow more exposure to the operators. However, without enough knowledge on the services characteristics (e.g. service priority, data packet profile, urgency, etc.), it is challenging for the operators to make a proper prediction and guarantee of QoE/QoS with optimized resource allocation, thus leading to a waste of network resources. Secondly, the infrastructure vendors generally have their private implementations of resource control and user scheduling algorithms, to which the mobile operators generally don't have sufficient control. This means even if the mobile operators have accurate mastery of users' QoE/QoS requirements, the eventual fulfillment of these requirements may be beyond control of the operators, especially when the radio resources are not sufficient enough during traffic peak period.

Despite the above challenges, the wireless communication industry has been working toward the performance improvement of extended reality (XR) and multimedia services for many years. For example, Service and System Aspect (SA) working group on "Services" (SA1) in 3GPP has studied the service requirements and key performance indicators (KPIs) of low latency services, interactive services, and tactile communications in 5G [4]. SA2 ("System Architecture and Services") has) standardized new 5G quality of service identifiers (5QI) to support interactive services including XR, featured by low latency and high data rate [5]. SA4 ("Multimedia Codecs, Systems and Services") has been documenting relevant traffic characteristics and providing a survey of XR applications. Since the radio access network (RAN) side functions are extremely important to satisfy the end-to-end mobile applications, 3GPP has studied the traffic models, simulation and evaluation methodologies in Release 17 in RAN1 ("Physical layer"), and management and optimization of QoE in RAN3 ("Overall UTRAN architecture"). In Release 18, the study is extended to understanding key features (e.g. periodicity, jitter, delay) of XR services at the base station (BS), based on which the QoS parameters and resource allocation will be optimized.

The above studies have been focused on better understanding and characterizing the mobile applications, and defining more accurate QoS parameters. However, these works have been carried out under the framework of 5G network architecture, to which large modification or evolution are not allowed in the 5G era. There exists an inherent gap between the mobile applications and the networks, especially when the applications need to adjust their data format to better match the fluctuations of wireless channels, because the applications generally don't have access to the related RAN information in time.

In order to resolve this contradiction, RAN capability exposure to the application layer has been investigated in the Internet Engineering Task Force (IETF) Application-Layer Traffic Optimization (ALTO) and Network Exposure Function (NEF), but with limited generality and flexibility. Mobile and Wireless Information Exposure (MoWIE), a framework to provide systematic information from networks to applications, was recently proposed [6]. However, there still seems a clear lack of tight coordination between the applications and networks. In particular, the applications may not know how its data would be eventually scheduled at the BS, and thus are unable to predict whether the QoE/QoS is satisfactory or not.

Toward a tight coordination or even convergence of applications and mobile networks in future 6G and beyond, a paradigm shift of the network architectures design is highly motivated. The structure and contribution of the paper are as follows. The design methodologies are discussed in Section II. Then, a multi-domain coordination scheme is proposed, where the predicted QoE/QoS at RAN is negotiated with the core network (CN) and application layer. A network architecture with cross-domain coordination center (CDCC) is further proposed, which collects and jointly processes the data from all the related domains and outputs control information to all the domains. In Section III, the benefits of the QoE assisted scheduling are investigated, where better throughput and fairness are demonstrated via simulations. In Section IV some future research directions and their standardization impacts are identified. The paper is concluded in Section V.

## II. Network Architecture Design toward Convergence of Applications and Networks

### A. Design Methodologies

The basic operation of wireless service provision is that the source-encoded service data from application layer will first go through the transport layer, then arrive at the CN, where the QoS parameters will be configured. Based on these QoS requirements, the real-time scheduler in the RAN will allocate proper wireless resources and configure suitable air interface technologies to transmit and receive the data packets. The CN will provide information to system operation, administration and management (OAM), which will be responsible for radio resource allocation between different BS or each network slice accordingly.

Essentially the QoS is an end-to-end service requirement, whose definition and provision should involve the application layer, transport layer, RAN, CN, mobile users, and even the OAM. However, this is not supported by the current network design methodology. Toward the end-to-end mobile service enhancements and a better tradeoff between fairness and efficiency, the interaction between the application layer and the network need be strengthened.

One approach to facilitating application and network convergence is conveying the high-dimensional application information to the network, including user experience indicators like QoE, and user profile, etc. These information traditionally are not available to the wireless networks and may have direct and important impacts on the system optimization. QoE is an important representative of the high-dimensional information, which can more intuitively and comprehensively evaluate the end-to-end service provision. Either predicted in the network or fed back from UE, QoE is able to facilitate a closed-loop wireless service provision. Additionally, if more accurate requirements of the services, especially those composed of multiple sub-services with distinguished features can be known to the CN/RAN, more efficient and reasonable resource allocation can be achieved.

Another approach is more exposure of the network information to the applications, rather than just the RAN capability. This information may include the QoE/QoS performance predicted by RAN, the (statistical) buffer status for each UE, the scheduling results, the statistical channel condition, etc. Based on these information, the application layer and the transport layer may more efficiently adjust their schemes.

Ideally, the above two approaches need to be jointly taken into consideration when designing the future network architecture. In the following, we will present two designs toward better convergence of applications and networks.

### B. Multi-domain Coordination Scheme

As the first step forward, a multi-domain coordination scheme is proposed in Fig.1, with the following characteristics:

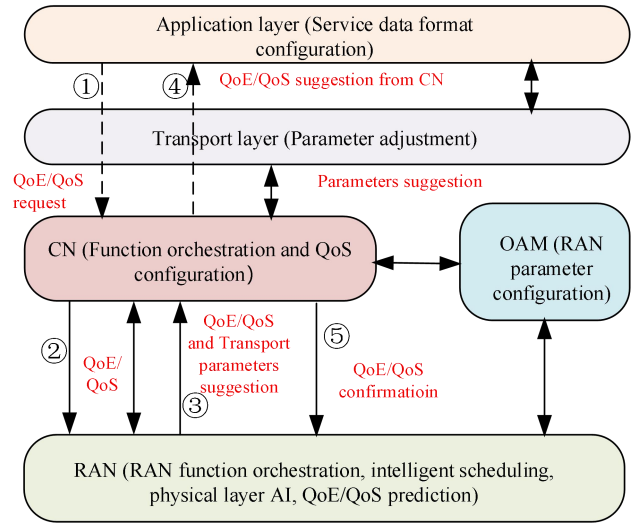

Fig.1 Architecture with multi-domain coordination

1) Based on the QoE/QoS requests from the applications, the CN is able to better understand the preferred QoE/QoS of each user or service slice and further determine the QoE/QoS parameters to improve on the fairness and efficiency. These QoE/QoS parameters will be transmitted to the RAN and set as performance targets during the user scheduling and wireless resource allocation.

Another approach for the RAN to have access to the QoE information is that the user equipment (UE) estimates or predicts the QoE performance in the application layer and then feeds it back to the RAN in case the user experience is not satisfactory. Based on this QoE information, the RAN is able to make a better suggestion to the CN about QoE/QoS.

2) The achievable QoE/QoS of each user guaranteed by the RAN scheduler can be predicted in the RAN, based on the analysis of users' service requirements and the corresponding channel conditions. The achievable QoE/QoS will be fed back to the CN to refine the QoE/QoS parameters. The final QoE/QoS, decided at the CN, will be used at the RAN. Via this closed-loop design, QoE/QoS is not merely dependent on the services but also dependent on RAN capabilities and other users' service requirements and channel conditions. Besides, the RAN can also predict possible transport layer parameter adjustment suggestion and feed it back to the transport layer.

3) In case the wireless network (CN and RAN) can not satisfy the mobile service requirements, the CN will inform the application server about the QoE/QoS suggestion for each user, and the application server will configure its source coding format (e.g. video definition and frame rate). With AI capabilities at the application server, the AI-enabled source coding optimization is also possible. For example, if the field of view can be predicted in the virtual reality system, the source coding can be focused on the predicted area and thus the source coding and transmission can be much simpler [7]. Another example is that the videos captured in the surveillance system can be analyzed by the AI capability and only the analysis results (either picture or mere text) are transmitted.

Intrinsically, the probability that the RAN fails to satisfy each user's service requirements could be significantly reduced following the design approach in Fig. 1. The involved interactions are multi-dimensional, between CN and RAN, CN and applications, applications and UEs, and even joint operation of CN, RAN, OAM, application, transport, and UEs. This ensures a better fairness between users and services, and better efficiency of network resource utilization. Note that for latency sensitive services the negotiation process need to be minimized.

*C. Network Architecture with CDCC*

The scheme in Fig.1 serves as an starting point for the network architecture evolution toward better convergence of applications and networks. The drawback of this scheme is that each domain doesn't have enough and real time information of other domains and thus the final decision making is not optimal. Besides, the negotiation between different domains may lead to unsatisfactory latency. A further step forward would be implementing the CDCC in the network, which functions as the mind of the whole network, and is responsible for information acquisition from all the domains, data processing, AI model training, inference, and control information generation for each domain. The logical architecture and the interfaces are shown in Fig.2, and the practical deployment of the CDCC can be in the third party platform, the CN, or the RAN.

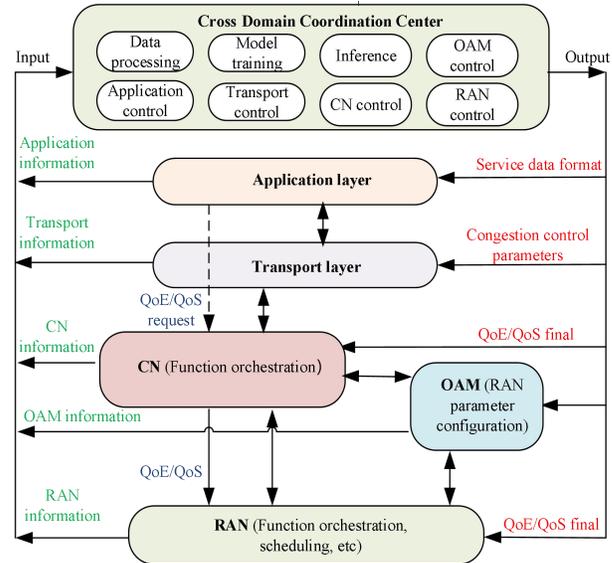

Fig.2  The Logical Network Architecture with CDCC

Different from the coordination scheme illustrated in Fig.1, with the joint processing of all the related information in the CDCC, the control information to all the domains can be globally optimized. For example, the service data format will be input to the application layer, based on which the source coding will be re-configured.

Traditionally, the transport layer like Transmission Control Protocol (TCP) cannot distinguish whether a drop is because of congestion or other flaws in the wireless communication. Most of the time, the congestion control mechanism will reduce the sending rate unnecessarily, leading to performance degradation. Enabled by the new architecture, the transport layer parameters (e.g. for the congestion control) will be input to the transport layer, which can more accurately track the radio link quality and improve the transport layer throughput.

The achievable QoE/QoS parameters will be calculated in the CDCC based on the the scheduling results (which user is scheduled and how many data is transmitted) from RAN and other information input from other domains. This information will be transmitted to the CN and RAN, which will be utilized to guarantee the QoS flow quality in CN and MAC scheduling in RAN.

The OAM control information can also input to the OAM, which will adjust its resource allocation strategy accordingly. For example, when the QoE/QoS is still not satisfactory after rounds of MAC scheduling, the OAM may allocate more resources to this cell via load balancing. In the extreme case where even the intelligent load balancing can not help, the CDCC need to collaborate with applications like navigation to change the navigation route if the UE is on a vehicle.

The time granularity of parameter configuration in different domains are quite diversified, e.g. the application layer data format is changed per seconds, the RAN function configuration and RAN resource management is conducted over one hundred milliseconds, while the user scheduling and resource allocation in the media access control (MAC) layer is per millisecond. Therefore, there are some practical limitations on the data collection from each domain. For lack of space, the design of the joint information processing in CDCC is not discussed in this paper, which is actually very complicated.

*C.1 Deployment of CDCC in the Third Party Platform*

When the CDCC is deployed in the third party platform, it may be difficult to collect real time data or information from all the domains, possibly with latency being over several hundreds of micro seconds. Therefore, the CDCC is not able to handle the real time inference or time sensitive control signaling, particularly for the RAN. For fast coordination between services and network, the multi-domain coordination scheme in Fig.1 may also be necessary in Fig.2. The key benefits of CDCC lie in its capability of global optimization of non-time-sensitive parameters. For example, based on the characteristics of the mobile applications, the CDCC is able to optimize the cache policy for the mobile service contents, long term resource allocation policy for each BS, the long term QoE/QoS targets of each user.

*C.2 Deployment of CDCC in the CN*

When the CDCC is deployed in the CN, data collection from the application layer, transport layer and the CN is simpler than in that in *C*.1. The initial QoE/QoS suggestion needs to be calculated in the RAN, because it would be challenging to feed back all the necessary RAN information to the CN in real time. The RAN makes fast prediction about the achievable QoE/QoS of each user based on information like the wireless channel conditions, service load, and buffer size and feeds back the statistical information to the CDCC, which will make final decision about QoE/QoS.

*C.3 Deployment of CDCC in the RAN*

When the CDCC is deployed in the RAN, it is able to collect all the necessary information from RAN like the MAC scheduler results, thus being able to make globally optimized prediction of the achievable QoE/QoS in the real time manner.

## III. QoE Information Assisted Scheduling: A Bridge to Application and Network Convergence

QoS has long been considered as the most successful system quality strategy that objectively measures the performance of a video transmission system from the source to the destination. However, QoS doesn't include the influence factors related to the physical, emotional and mental constitution of the mobile users and context information of the social and economical background and status. Consequently, QoS is unable to reflect accurately the true experience of XR services in the mobile communication networks.

Thanks to its advantages, QoE has been widely recognized as a preferred strategy in the context of video and XR transmissions. QoE management schemes have been surveyed in [2], where QoE optimization of multimedia services is implemented in the CN or mobile edge, which only indirectly affects the resource allocation of RAN. With the architectures proposed in Section II.B, the QoE of each user is jointly determined based on the information input from all the related domains, whose reliability and accuracy is greatly improved. In this section, we assume the CDCC is deployed in RAN and therefore is capable of inference on QoE and QoS in real-time or near real-time manner (milli-second level). We will show the benefits brought by the QoE information assisted MAC scheduling in the RAN.

*A. MAC Scheduler Enabled by the New Network Architecture*

Different from traditional MAC schedulers which generally utilize the QoS information (e.g. the latency and packet loss rate requirement) to allocate the wireless resources to each user, the proposed MAC scheduler BCQQ (Buffer, CQI, QoS, and QoE) utilizes the buffer condition, channel quality indicator (CQI), QoS parameters, and QoE information as input, as shown in Fig. 3.

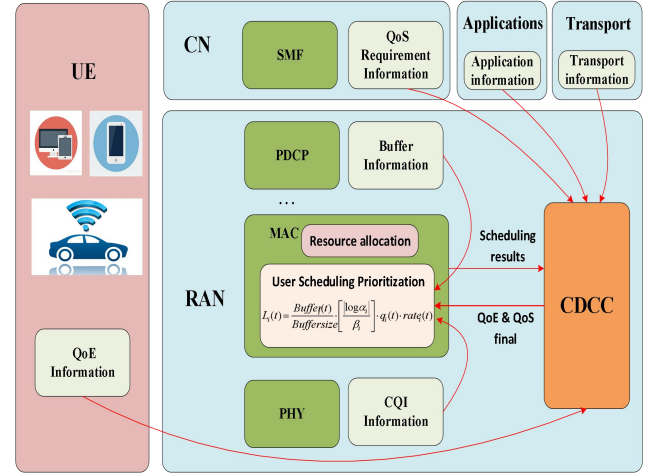

Fig. 3. The MAC scheduler enabled by the new network architecture

During data collection period, the buffer information, especially its dynamic change, can be collected by the interface between Packet Data Convergence Protocol (PDCP) layer and MAC layer. The collection period is consistent with scheduling period, generally at millisecond level. CQI information is directly collected by the physical layer feedback. QoS requirement information like latency and packet loss rate requirement can be directly configured by the session management function (SMF) in CN, and utilized as input to the CDCC. The QoE information is calculated by the UE in the application layer and fed back to the BS through a new interface between UE and the BS. This QoE information goes directly from the application layer to the RAN and is utilized in the CDCC, which will generate the final QoE and QoS based on all the information input. The final QoE and QoS will serve as the performance target for the MAC scheduling.

Simply put, the basic principal of the QoE assisted scheduler is that for each transmit slot one user with the highest scheduling priority will be selected. In calculating the priority of the *i*-th user, the following equation is adopted.

$$L_i(t) = \frac{Buffer_i(t)}{Buffersize} \cdot \left[\frac{|\log \alpha_i|}{\beta_i}\right] \cdot q_i(t) \cdot rate_i(t) \quad (1)$$

In Eq.1, *Buffersize* represents the total buffer space allocated by the BS for each user. $Buffer_i(t)$ represents the occupied buffer at time *t*. When the occupation ratio gets higher, the priority of this user need to increase. *α* and *β* represent target packet loss rate and delay of the traffic, respectively. The priority increases with a lower *α* and lower *β*. $q_i(t)$ is the QoE for the *i*-th user, representing the target satisfaction degree of the mobile service requirements. Note that there are various modeling and evaluation methods regarding the QoE, how the method is selected is not discussed in this paper. The achievable $rate_i(t)$ is calculated based on the CQI feedback, which represents the amount of data that the channel can support for the *i*-th user at time *t*. The designed scheduler algorithm has the following features.

a) The algorithm takes into consideration the buffer occupation ratio, which represents the matching degree between the capability of air interface to send data and service arrival data volume. A higher buffer occupation ratio leads to a higher priority. Generally the buffer size for each user is limited. In case the buffer occupation ratio is high and the scheduling priority is still comparatively low, the CDCC need to send service adjustment request to the application layer and transport layer, which will reduce the service data volume accordingly. This can help to significantly avoid the overflow of the buffer. Alternatively, the BS can allocate dynamic buffer size for each user, to alleviate the burden on application layer and transport layer. The dynamic partition of memory size for data buffering and data processing is also an interesting research topic.

b) The essential QoS parameters like packet loss rate and delay are included in the algorithm, which are designed to satisfy the service requirements in the wireless communication network. The stricter the requirements on delay or packet loss rate, the higher the scheduling priority. However, traditional scheduling policies to meet the requirement of these QoS parameters will not necessarily lead to satisfactory QoE performance. QoE oriented scheduling is fundamentally motivated, where the QoE information serves as an efficient metric to further refine the QoS fulfillment scheduling policy.

c) The buffer occupation ratio, packet loss, service delay, achievable air interface rate and QoE information are jointly considered. The real-time scheduler output (e.g. how much resource is allocated and how many data is transmitted for each user) will be input to the CDCC, which will make intelligent decisions on the application layer format, transport layer scheme, buffer size configuration, and QoE/QoS target for each user to maximize the end-to-end mobile services. This is not possible in the current 5G network architecture. For example, the scheduler can predict the buffer status in the future time window, and this information can be input to the transport layer, which will accordingly adjust its data transmit policy, to better track the wireless channel and the buffer status. Another example is the source coding adjustment in the application layer. With the global optimization capability in the CDCC, the source coding of all the users' applications can be jointly configured, which can match the transport layer capability and the air interface capability, to maximize the end-to end QoE/QoS performance.

B. *Performance Evaluation*

The following simulation is given to show the performance of the proposed scheduler, with simulation parameters listed in Table I.

TABLE I. Simulation Parameters

| User number | 5 | Cell number | 1 |
|---|---|---|---|
| Traffic I | FTP Download | BS number | 1 |
| Traffic II | Live HD Video | Cell radius | 1km |
| Cell peak rate | 6Gbps | Moving speed | 3km/h |
| Buffersize | 5MB/UE | Scheduling period | 1ms |

Five users (No.1 to No.5) are considered. The traffic type of No.1-No.3 users is FTP (File Transfer Protocol) download, with a 10^-6 packet loss rate, an acceptable delay of 300ms, and a 0.5Mb average download traffic packet size. The traffic type of other users (No.4-No.5) is Live HD Video, with a 10^-6 packet loss rate, an acceptable delay of 150ms, a maximum packet size of 2Mb, and random packet arrival distribution.

To show the performance improvement from the above proposed QoE assisted BCQQ scheduling, we choose the classic scheduler MLWDF [1] and LEASCH [8] as references. MLWDF is an optimization of proportional fairness algorithm, while LEASCH is a pure deep reinforcement learning algorithm for radio resource scheduling in 5G MAC.

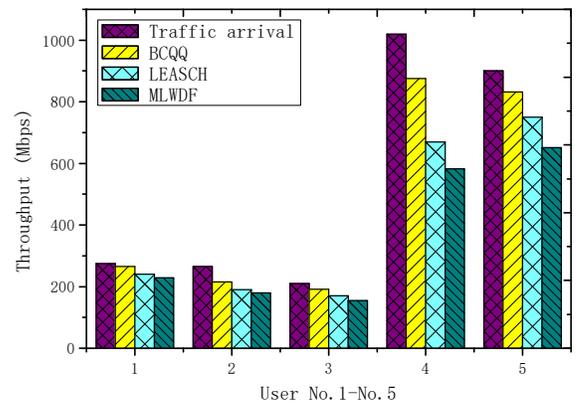

Fig. 4. Performance in multi-traffic scenario

Fig.4 shows the throughput performance comparison of all the three schedulers. We can see that the arrived data volume of Live HD Video far exceeds that of FTP download. The simulation results show the overall system throughput of all the five users are 2378 Mbps with BCQQ, 2019 Mbps with LEASCH, and 1796 Mbps with MLWDL The performance gains of BCQQ over LEASCH and MLWDL are 17.8% and 32.4%, respectively. Compared with LEASCH and MLWDL, QoE-assisted BCQQ scheduler allocates more scheduling opportunities to Live HD Video users based on QoE feedback while ensuring the QoS requirements of FTP users.

*C. QoE Oriented Fairness Index*

Jain's fairness index (JFI) [9] has been widely utilized to quantify fairness among various fairness measures due to its highly intuitive interpretation. During a certain time window, the closer is the actual amount of each user's data has been scheduled, the better is the fairness, with JFI approaching to 1 when the scheduled data size is same for all users. However, JFI becomes inappropriate in scenarios with diversified service types, such as Live HD, game, voice, and instant messaging. The experience of high data rate applications like Live HD would be seriously frustrated if only absolute volume of service data is considered. To better reflect the QoE/QoS satisfaction of the scheduled users, in the following, we design a new fairness index based on users' QoE.

$$QoE\_FI = \sum_{i=1}^{n}\sum_{j=1}^{n}\left|\frac{y_i}{Y_{QoE-i}} - \frac{y_j}{Y_{QoE-j}}\right|, \; i \neq j \quad (2)$$

In Eq.2, $QoE\_FI$ is the fairness index within a period of time, which can be multiple scheduling time slots. $n$ is the active user number scheduled in the BS, $y_i$ represents the amount of traffic data of the *i-th* user which has been scheduled and $Y_{QoE-i}$ represents the data volume which need to be sent via the air interface to guarantee the QoE performance. $y_i/Y_{QoE-i}$ represents the QoE satisfaction ratio and need to be maintained similar among all the users via a fair scheduler. The smaller value of *FI*, the better fairness.

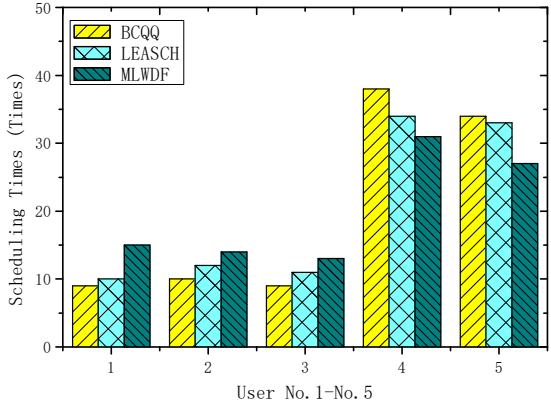

Fig. 5. Scheduling times comparison in multi-traffic scenario

Fig. 5 shows the comparison of the number of scheduling times in the simulation in III.B. Because this is a multi-traffic scenario, the JFI and our proposed fairness index have obvious difference. The JFIs are 0.71, 0.72 and 0.74 with BCQQ, LEASCH, and MLWDL scheduler, respectively, which are very close, with the BCQQ JFI being slightly smaller than that of the other two schedulers. But as shown in Fig. 4, there is a large improvement in throughput performance with the proposed BCQQ scheduler. This indicates substantial performance gains are possible with BCQQ scheduler with a neglegiable JFI loss.

Since the JFI can hardly reflect the QoE satisfaction of each user, it is not sensitive and appropriate enough to judge fairness for the multi-traffic scenarios. Taking the proposed $QoE\_FI$ as the metric, the fairness index is 0.74, 1.08 and 1.15 with BCQQ, LEASCH, and MLWDL respectively, the proposed BCQQ scheduler having the best fairness. Larger gaps in the fairness are observed in these three schedulers, indicating the $QoE\_FI$ is more sensitive and effective for multi-traffic scenarios.

Traditionally, the efficiency (e.g. the throughput) and fairness of the scheduler are contradictive, a higher throughput generally leading to a worse fairness. The proposed QoE assisted BCQQ scheduler has exhibited better throughput performance and better fairness index with $QoE\_FI$ metric, showing an enhanced efficiency-fairness tradeoff.

## IV. FUTURE RESEARCH DIRECTIONS AND STANDARDIZATION IMPACTS

The approaches presented in Section II bring a paradigm shift of the network design, necessitating tremendous research and standardization studies.

For the multi-domain coordination scheme, the coordination schemes need to be studied and specified.

1) How can the AI algorithms in CN or RAN accurately predict achievable QoE/QoS for each service and what are the signaling interactions between CN and RAN, and between CN and applications, regarding the negotiation of QoE/QoS parameters? This is vastly different from current mechanisms specified in 3GPP.

2) If it is the mobile user who calculates the QoE information in the application layer and feeds it back to the BS, the corresponding signaling interactions between UE and BS also need to be specified.

For the architecture with CDCC, there seems a long way to reach the final target. The following issues need to be resolved.

1) The first challenge would be the data collection from all the related domains. The data from applications generally belong to the over-the-top service providers, while the wireless network data belong to the mobile operators, infrastructure vendors, and mobile terminal vendors. Collection and joint processing of these data require more technical, managerial, and law support to ensure the data security.

2) How to specify the cross-domain interfaces (e.g. the signaling interactions and data formats) requires joint efforts from all the domains and all the related standardization bodies like IETF, 3GPP SA, CN, and RAN.

3) How the application layer data processing like source coding evolves to integrate the information of transport layer,

4CN and RAN to ensure the best end-to-end QoE/QoS performance need further investigations and specifications. In essence, the design issue is how to encode the source optimally if the end-to-end data path is known, which is similar to the joint source-channel joint encoding.

4) How to further optimize the transport layer to better match the transport schemes with the source coding and radio link fluctuations is also a very important research and standardization direction [10].

5) How to jointly optimize the parameters from different domains in CDCC is fundamentally a challenging task. This may require a super AI model like GPT-4, being in charge of all the optimization issues.

6) The current research focus of AI enabled RAN architecture is how to introduce AI platform and functions in BSs to enhance the performance of traditional radio resource management, radio resource control, and user scheduling schemes. This will promote openness of the existing 4G and 5G BSs. However, these "add on" AI features were not meant to change the air interface communication protocols. To fully unleash the potential of AI capabilities in wireless communications, the quest for AI enabled design of air interface for 5G [11] and future 6G communication networks [12] is well motivated.

## V. Conclusions

In this paper, we have presented a design framework of convergence of mobile applications and networks. The challenges faced with the industry is first analyzed. The design methodologies are further discussed, following by the design of a multi-domain coordination scheme and a network architecture with cross-domain coordination center. The benefits of the QoE assisted scheduling are investigated, where better throughput and fairness are demonstrated via simulations. Some future research directions and their standardization impacts are identified.

## References

[1] M. Andrews, et al., "Providing quality of service over a shared wireless link," *IEEE Commun. Mag.*, vol. 39, no. 2, pp. 150–154, Feb. 2001.

[2] A. A. Barakabitze et al., "QoE management of multimedia streaming services in future networks: A tutorial and survey," *IEEE Commun. Surveys & Tuts.*, vol. 22, no. 1, pp. 526-565, 1st Quarter 2020.

[3] Y. Chen, K. Wu, and Q. Zhang, "From QoS to QoE: A tutorial on video quality assessment," IEEE Commun. Surveys & Tuts., vol. 17, no. 2, pp. 1126–1165, 2nd Quarter 2015.

[4] 3GPP, TS 22.261 "Service requirements for the 5G system," v.18.5.0, Dec. 2021.

[5] 3GPP, TS 23.501 "System architecture for the 5G System (5GS)", v.17.3.0, Dec. 2021.

[6] Y.F. Zhang, et al. "MoWIE: Toward Systematic, Adaptive Network Information Exposure as an Enabling Technique for Cloud-Based Applications over 5G and Beyond," *ACM Sigcom* 2020, Aug. 2020.

[7] D. B. Kurka and D. Gündüz, "DeepJSCC-f: Deep joint source-channel coding of images with feedback," *IEEE J. Sel. Areas Inf. Theory*, vol. 1, no. 1, pp. 178-193, May 2020.

[8] F. Al-Tam, N. Correia, and J. Rodriguez, "Learn to schedule (LEASCH): A deep reinforcement learning approach for radio resource scheduling in the 5G MAC layer," *IEEE Access*, vol. 8, pp. 108088–108101, 2020.

[9] R. Jain, D. Chiu, and W. Hawe, "A quantitative measure of fairness and discrimination for resource allocation in shared systems," Digital Equipment Corporation, Tech. Rep. DEC-TR-301, Sep. 1984.

[10] S. R. Pokhrel and M. Mandjes, "Improving multipath TCP performance over WiFi and cellular networks: An analytical approach," IEEE Trans. Mobile Comput., vol. 18, no. 11, pp. 2562-2576, Nov. 2019.

[11] RP-213599, New SI: Study on Artificial Intelligence (AI)/Machine Learning (ML) for NR Air Interface.

[12] S. Han et al. "Artificial intelligence enabled air interface for 6G: solutions, challenges, and standardization impacts," *IEEE Commun Mag*. vol.58, no.10, pp.73-79, 2020.

## Biographies

Shuangfeng han (hanshuangfeng@chinamobile.com) received his Ph.D. degree in Electronic Engineering from Tsinghua University in 2006. He is a Principal Researcher of China Mobile Research Institute. His research interests include massive MIMO, NOMA, and wireless artificial intelligence. He is the recipient of IEEE Comsoc 2018 Fred W. Ellersick Award and IEEE Comsoc AP Outstanding paper award. He is an IEEE Senior Member, and is serving on the editorial board of IEEE communications magazine.

Zhiming Liu (liuzhiming@chinamobile.com) received the M.S degree in electronic and communication engineering from Beijing University of Posts and Telecommunications in 2016. He has been with China Mobile Research Institute after graduation. His research focuses on AI driven intelligent RAN optimization, network resource management, protocol stack and signaling design, etc.

Tao Sun (suntao@chinamobile.com), IEEE member, received the B.S. degree with the Department of Automation, in 2003 and Ph.D. degree in Control Science and Engineering, in 2008, both from Tsinghua University, Beijing, China. He jointed China Mobile Research Institute from 2008, serving as Chief Expert since 2021. He was rapporteurs of many 3GPP work items such as Study of Next Generation Mobile Architecture, 5G System Architecture Phase 1, Enhancement of Service-based Architecture. He now is the Vice Chair of 3GPP SA2 and China Mobile's coordinator of 3GPP SA and CT groups.

Xiaoyun Wang (wangxiaoyun@chinamobile.com) is the general manager of Technology Department of China Mobile. Her research interests include technology strategy, system architecture and networking technology. She is the recipient of multiple National Science and Technology Progress Awards, the National Innovation and Excellence Award, and the Chinese Youth Science and Technology Award.